\documentclass{article}
\usepackage{spconf,amsmath,graphicx}
\ninept
\usepackage{amsfonts}
\usepackage{amssymb}
\usepackage{relsize,exscale}
\usepackage{color}%
\usepackage{array}
\usepackage{subfig} 

\graphicspath{{Figures/}}

\newcommand{\bv}[1]{\boldsymbol{#1}}
\newcommand{\unit}[1]{\bv{\hat{#1}}}

\newcommand{\sglt}[1]{\mathfrak{S}_{\rm GL}^T(#1)}
\newcommand{\sglq}[1]{\mathfrak{S}_{\rm GL}^Q(#1)}

\newcommand{\untsph}{{\mathbb{S}^{2}}} 

\newcommand{\secref}[1]{Section\,\ref{#1}} 
\newcommand{\figref}[1]{Fig.\,\ref{#1}}
\newcommand{\lsph}{L^{2}(\mathbb{S}^{2})}

\newcommand{\sseq}[1]{\mathfrak{S}_{\rm E}^Q(#1)}
\newcommand{\sset}[1]{\mathfrak{S}_{\rm E}^T(#1)}

\newcommand{\sso}[1]{\mathfrak{S}_{\rm O}(#1)}

\title{Sampling Schemes for Accurate Reconstruction and Computation of Performance Parameters of Antenna Radiation Pattern}
\name{Umair Ahmed and  Zubair~Khalid , {Member,~IEEE}\thanks{Z.~Khalid is supported by Pakistan
			HEC 2016-17 NRPU (Project no. 5925)}}

\address{
     School of Science and Engineering, Lahore University of Management Sciences, Lahore, Pakistan\\
Email:
umair.ahmed@lums.edu.pk, zubair.khalid@lums.edu.pk}

\begin{document}
%
\maketitle
\vspace{-1mm}
\begin{abstract}
In practice, the finite number of samples of the spherical radiation pattern or antenna gain are taken on the sphere for both the reconstruction of the antenna radiation pattern and the computation of mobile handset performance measures such as directivity and mean effective gain~(MEG). The acquisition of samples is time consuming as the measurements are required to be collected over the range of frequencies and in multiple spatial directions. It is therefore desired to have a sampling strategy that takes fewer number of samples for the accurate reconstruction of radiation pattern and incoming signal power distribution. In this work, we propose to use equiangular sampling, Gauss-Legendre sampling and optimal dimensionality sampling schemes on the sphere for the acquisition of measurements of spherical radiation pattern of the antenna for its reconstruction, analysis and evaluation of performance parameters of the antenna. By appropriately choosing the spherical harmonic degree band-limits of the gain and the power distribution model of the incoming signal, we demonstrate that the proposed sampling strategies require significantly less number of samples for the accurate evaluation of MEG than the existing methods that rely on the approximate evaluation of the surface integral on the sphere.
\end{abstract}

\vspace{-1mm}

\vspace{-1mm}
\section{Introduction}
\label{sec:intro}

In the design and analysis of antenna systems, it is important to characterize the radiation pattern of antennas and their performance parameters~\cite{carro2010radiation, misra2006design}. Antenna radiation patterns serve to visually understand the radiation characteristics of an antenna under experimentation and are crucial to understand the geometric dependence of radio waves from the antenna. To parameterize the quality of an antenna, radiation patterns are used to derive other measures such as mean effective gain (MEG) and antenna directivity. Hence, these radiation patterns form the basis of design and analysis of antenna systems.

Instead of near-field characterization that are costly and difficult to implement in practice for large range of frequencies, the far field measurements are taken for antenna radiation pattern reconstruction and evaluation of performance parameters. Besides their practical advantages, the far field characterization requires large number of measurements taken at different frequencies on the sphere in multiple directions and for different power distributions of incoming signal~\cite{nielsen2006mobile}. Antenna radiation patterns are represented as a function of azimuth and elevation angle on the sphere. For theoretical analysis, the antenna patterns are usually obtained by simulations carried out using finite difference time domain~(FDTD)~\cite{tirkas1992finite} and Method of Moments (MoM)~\cite{sarkar2000method} techniques. However under experimental conditions, the strength of the radio waves are measured by means of a probe antenna and successively rotating the antenna in fixed increments over the whole range of azimuthal and elevation angles~\cite{nielsen2006mobile}. These data-points are then used to reconstruct the antenna radiation pattern. Since the acquisition set-up is costly and time consuming, it is of significant importance to adopt a sampling scheme for the acquisition of measurements that enable accurate and fast reconstruction of antenna radiation pattern.

Sampling the antenna radiations over all possible combinations of azimuthal and elevation angles significantly increases the time-cost of the whole process. To address this issue previously, \cite{nielsen2006mobile} has worked on optimizing the amount of fixed increment in the two angles to meet an acceptable trade-off between accuracy and time complexity. Similarly, \cite{flint2010low} has used the famous sphere partitioning algorithm by \cite{leopardi2006partition} to reduce the resolution of spherical sampling for reconstructing the radiation patterns. The overall goal of all of the current techniques is to minimize the number of data points needed for an antenna analysis.

In this work, we propose to use different sampling schemes developed for the band-limited signals to sample the antenna radiation patterns and incoming signal power distribution, approximated as band-limited signals on the sphere~\cite{bucci1987spatial}. We consider the iso-latitude sampling schemes such as equiangular~\cite{McEwen:2011}, Gauss-Legendre~\cite{Doroshkevich:2005} and optimal-dimensionality~\cite{Khalid:2014} as these schemes enable accurate signal reconstruction supported by the computation of spherical harmonic transform of band-limited signals with reconstruction error on the order of numerical precision. We also develop formulation to determine the band-limited approximation of the radiation pattern for different sampling schemes. We also present variants of the sampling scheme for the computation of quadrature involved in the evaluation of performance parameters of the antenna radiation pattern. Using the proposed formulation, we analyse the band-limit of radiation pattern and incoming signal power distribution and demonstrate that the use of proposed sampling scheme requires less number of samples than those required by existing methods.

\vspace{-1mm}
\section{Preliminaries and Problem formulation}

\subsection{Gain or Antenna Radiation Pattern as Signal on the Sphere}
The radiation pattern or gain pattern or simply gain of the antenna for fixed frequency $f$ can be represented as a signal on the unit sphere $\untsph$, of the form\footnote{It is considered that the radiation pattern depends on the frequency. We drop this dependence for notational convenience.} $G(\unit{x};f)\equiv G(\unit{x})$, where $\unit{x}\equiv \unit{x}(\theta, \phi) \triangleq (\sin\theta\,\cos\phi, \; \sin\theta\,\sin\phi, \; \cos\theta)$ represents a point in the 3D space on the unit sphere. Here $\theta \in [0,\,\pi]$ is the co-latitude angle, $\phi \in [0, \, 2\pi)$ is the longitude angle and $(\cdot)^{\mathsf{T}}$ denotes the transpose operator. We assume that the antenna radiation pattern $G(\theta,\phi)$ lies in the Hilbert space, denoted by $L^2(\mathbb{S}^2)$, of square integrable functions on the sphere. Hilbert space $L^2(\mathbb{S}^2)$ is equipped with the inner product defined for two functions $F,G \in L^2(\mathbb{S}^2)$ as

\begin{equation}
\label{eqn:innprd}
\langle F,G \rangle = \int_{\untsph} F(\unit{x})G^*(\unit{x})ds(\unit{x}),
\end{equation}
where $ds(\unit{x}) = \sin(\theta) d\theta d\phi$ is the differential area element, $(\cdot)^{*}$ represents the complex conjugate operation and the integration is carried out over the whole sphere, that is, $\displaystyle \int_{\untsph} = \displaystyle \int_{\theta=0}^\pi \displaystyle \int_{\phi=0}^{2\pi}$. The inner product defined in (\ref{eqn:innprd}) induces a norm $\|G\| \triangleq\langle G,G \rangle^{1/2}$. The energy of the function $G$ is given by $\|G\|^2$.

\subsection{Antenna Performance Parameters}
\label{sec:parameetrs}
To quantitatively define the performance of any antenna, an accurate representation of the antenna radiation pattern is of paramount importance. The mean effective gain (MEG) and directivity being the key representatives of the antenna quality are calculated directly from these radiations patterns. Directivity of an antenna signifies the extent to which, the shape of its radiation pattern is directional. A perfectly isotropic antenna has therefore unit directivity and for an antenna with radiation pattern $G(\theta,\phi)$, the directivity is given as
\begin{equation}
\label{eq:directivity}
D = \frac{G_{\textrm{max}}}{G_{\textrm{av}}} = \frac{\max\limits_{\unit{x}\in\untsph}  G(\unit{x})} {\displaystyle \frac{1}{4\pi}\int_\untsph G(\unit{x}) ds(\unit{x})}.
\end{equation}
The mean effective gain~(MEG) quantifies the antenna quality by incorporating the complete range of directional and polarization variations of the incoming signal power and the receiver and is given by~\cite{jakes1974microwave,taga1990analysis}
\begin{align}
\label{eq:gamma}
\Gamma  = \frac{\displaystyle \int_{\untsph}\big( G_\theta(\unit{x}) Q_\theta(\unit{x}) + G_\phi(\unit{x})Q_\phi(\unit{x})    \big) ds(\unit{x})  }{ \displaystyle \int_{\untsph} \big( Q_\theta(\unit{x}) + Q_\phi(\unit{x}) \big) ds(\unit{x}) }.
\end{align}
In \eqref{eq:gamma}, $G_\theta(\unit{x})$ and $G_\phi(\unit{x})$ correspond to the antenna power gain for $\theta$ and $\phi$ polarization respectively. Similarly, $Q_\theta(\unit{x})$ and $Q_\phi(\unit{x})$ denote the average power of the incoming signal incident on the antenna in the direction $\unit{x}$. The dependence of the incoming signal average power on the frequency is dropped in the adopted notation.

\subsection{Incoming Signal Power Model}
\label{sec:models}

To model either $Q_\theta(\unit{x})$ or $Q_\phi(\unit{x})$, we consider a real world power density model, referred to as HUT model~\cite{kalliola2002angular}, where the incoming power signal is modeled as azimuthally symmetric and distributed according to the double exponential distribution in elevation as\footnote{We note that we use $G(\unit{x})$ and $Q(\unit{x})$ to represent radiation pattern for both $\theta$ and $\phi$ polarization components of these signals.}
\begin{equation}
\label{Eq:hut}
Q(\unit{x}) = Q(\theta)=\begin{cases}
K \exp\big(-\frac{\sqrt{2}|\theta-\theta_o|}{\sigma^{-}}\big),& \theta\in[0,\theta_o],\\
K \exp\big(-\frac{\sqrt{2}|\theta-\theta_o|}{\sigma^{+}}\big),& \theta\in[\theta_o,\pi],
\end{cases}
\end{equation}
where $\theta_o$ represents the mean, $\sigma^{-}$ and $\sigma^{+}$ quantify the spread around the mean in different directions, $|\cdot|$ denotes the absolute value and $K$ is chosen that ensures $\|Q\|=1$.

\subsection{Problem Under Consideration -- Review}

In this work, we consider the problem of spherical sampling of the antenna radiation pattern for its accurate reconstruction and the computation of antenna performance parameters defined in \secref{sec:parameetrs} for incoming power model given in \secref{sec:models}. Before we present the sampling strategies proposed in this work for sampling antenna radiation pattern and incoming signal average power, we briefly review the existing sampling schemes adopted for taking measurements of the antenna radiation pattern. Antenna radiation patterns are currently acquired by exhaustively sampling the value of antenna radiation gain along the complete range of azimuthal $\theta$ $\in$ $[0,2\pi]$, and elevation angle $\phi$ $\in$ $[0,\pi]$. A motorized positioning device capable of rotation along the two axes is installed inside an anechoic room. The antenna under test~(AUT) is placed over the motorized setup and a GSM tester, acting as the base station, is used to measure the power radiated from the antenna at each pair of azimuthal and elevation angle \cite{nielsen2006mobile}. The effect of decimating the range of azimuthal and elevation angles has been studied by \cite{nielsen2006mobile} to evaluate a trade-off between accuracy and time-cost of acquiring the radiation pattern. Under a decimation factor of $p$ and $q$ for $\theta$ and $\phi$ respectively, the total sampling points $N_s$ are given as
\begin{equation}
N_s = N_\theta N_\phi
\end{equation}
\noindent where $N_\theta = \frac{\pi}{p}$ and $N_\phi = \frac{2\pi}{q}$. For example, a decimation factor of $1^\circ$ for both azimuthal and elevation angle adopted to evaluate antenna performance parameters in \cite{nielsen2006mobile} using $N_s = 64,800$ total sampling points. The directivity and MEG are computed by approximating the integrals given in \eqref{eq:directivity} and \eqref{eq:gamma} respectively as

\begin{equation}
D \simeq \frac{\max G(\theta,\phi)}{\frac{1}{N_s}\sum\limits_{i=0}^{N_\theta-1}\sum\limits_{j=0}^{N_\phi-1} G(\theta_i,\phi_j)},
\end{equation}
\begin{equation}
\label{Gamma_sampling_method}
\begin{split}
\Gamma \simeq \sum_{i=0}^{N_\theta-1} \sum_{j=0}^{N_\phi-1} [G_\theta(\theta_i,\phi_j)Q_\theta(\theta_i,\phi_j) + & \\ G_\phi(\theta_i,\phi_j)Q_\phi(\theta_i,\phi_j)] \frac{\sin(\theta_i)}{P_{env}},
\end{split}
\end{equation}

\noindent{where}

\begin{equation}
P_{env} = \sum_{i=0}^{N_\theta-1} \sum_{j=0}^{N_\phi-1} [Q_\theta(\theta_i,\phi_j) + Q_\phi(\theta_i,\phi_j)]\sin(\theta_n)
\end{equation}

\vspace{-1mm}
\section{Sampling Schemes for Radiation Pattern Measurements}

We here present different sampling schemes using which the antenna radiation pattern can be reconstructed and its directivity and MEG can be computed with control on the accuracy of reconstruction and computation respectively. In comparison with the existing sampling strategies adopted for sampling antenna radiation pattern, the schemes we propose to use require significantly less number of samples and enable accurate signal reconstruction. We first expand the radiation pattern $G(\unit{x})$ using spherical harmonic functions -- orthonormal complete basis for $\lsph$ as

\begin{equation}
\label{Eq:G_expansion}
G(\unit{x}) = \sum_{\ell=0}^{\infty} \sum_{m=-\ell}^{\ell} G_\ell^m\, Y_\ell^m(\unit{x}).
\end{equation}
Here, $Y_l^m$ is the spherical harmonic function defined as~\cite{Kennedy-book:2013}
\begin{equation}
Y_\ell^m(\unit{x}) \triangleq \sqrt{\frac{2\ell+1}{4\pi} \frac{(\ell-m)!}{(\ell+m)!}} P_\ell^m(\cos\theta) e^{im\phi},
\end{equation}
for degree $\ell\geq0$ and integer order $|m| \leq \ell$ and $P_\ell^m(.)$ is the asscoiated Legendre function of degree $\ell$ and order $m$. $G_\ell^m$ is the spherical harmonic coefficient of degree $\ell$ and order $|m|\le\ell$, forms the spectral~(or harmonic) domain of $G(\unit{x})$ and is defined by the spherical harmonic transform~(SHT) given by
\begin{equation}
\label{eq:sht}
G_\ell^m = \langle G,Y_\ell^m \rangle \triangleq \int_{\mathbb{S}^2} G(\unit{x}) \overline{Y_\ell^m(\unit{x})} ds(\unit{x}).
\end{equation}
We define the $\epsilon_o$-band-limited approximation of the radiation pattern $G(\unit{x})$ as $\tilde{G}(\unit{x})$ band-limited at spherical harmonic degree $L$ and is given by
\begin{equation}
\label{Eq:tildeG}
\tilde{G}(\unit{x}) = \sum_{\ell=0}^{L-1} \sum_{m=-\ell}^{\ell} G_\ell^m Y_\ell^m(\unit{x}),
\end{equation}
such that
\begin{align}
\label{Eq:error_bound}
E_G = \frac{||\tilde{G}(\unit{x}) - {G}(\unit{x})||}{|| {G}(\unit{x})||} < \epsilon_o.
\end{align}
We note that the incoming signal average power $Q(\unit{x})$ can also be similarly expanded in terms of spherical harmonic basis functions. In order to accurately reconstruct the radiation pattern with error of norm less than $\epsilon_o$, we are required to sample the radiation pattern such that the SHT given in \eqref{eq:sht} can be accurately computed. For the computation of performance parameters such as directivity and MEG, samples should be taken such that the integral of $G_\theta(\unit{x})$,  $G_\theta(\unit{x})Q_\theta(\unit{x})$ and $G_\phi(\unit{x})Q_\phi(\unit{x})$ can be computed. We now present different sampling schemes proposed in the literature and their variants which enable computation of SHT and integral of the signal band-limited at degree $L$ by taking $O(L^2)$ number of measurements. To facilitate the acquisition of measurements of radiation pattern, we focus on the iso-latitude spherical sampling schemes composed of iso-latitude annuli rings of samples that may or may not be uniformly spaced along co-latitude. However, the samples along $\phi$ in each ring are placed with equiangular spacing.

\begin{figure}[!t]
\centering
\vspace{-6mm}
\includegraphics[width=0.4\textwidth]{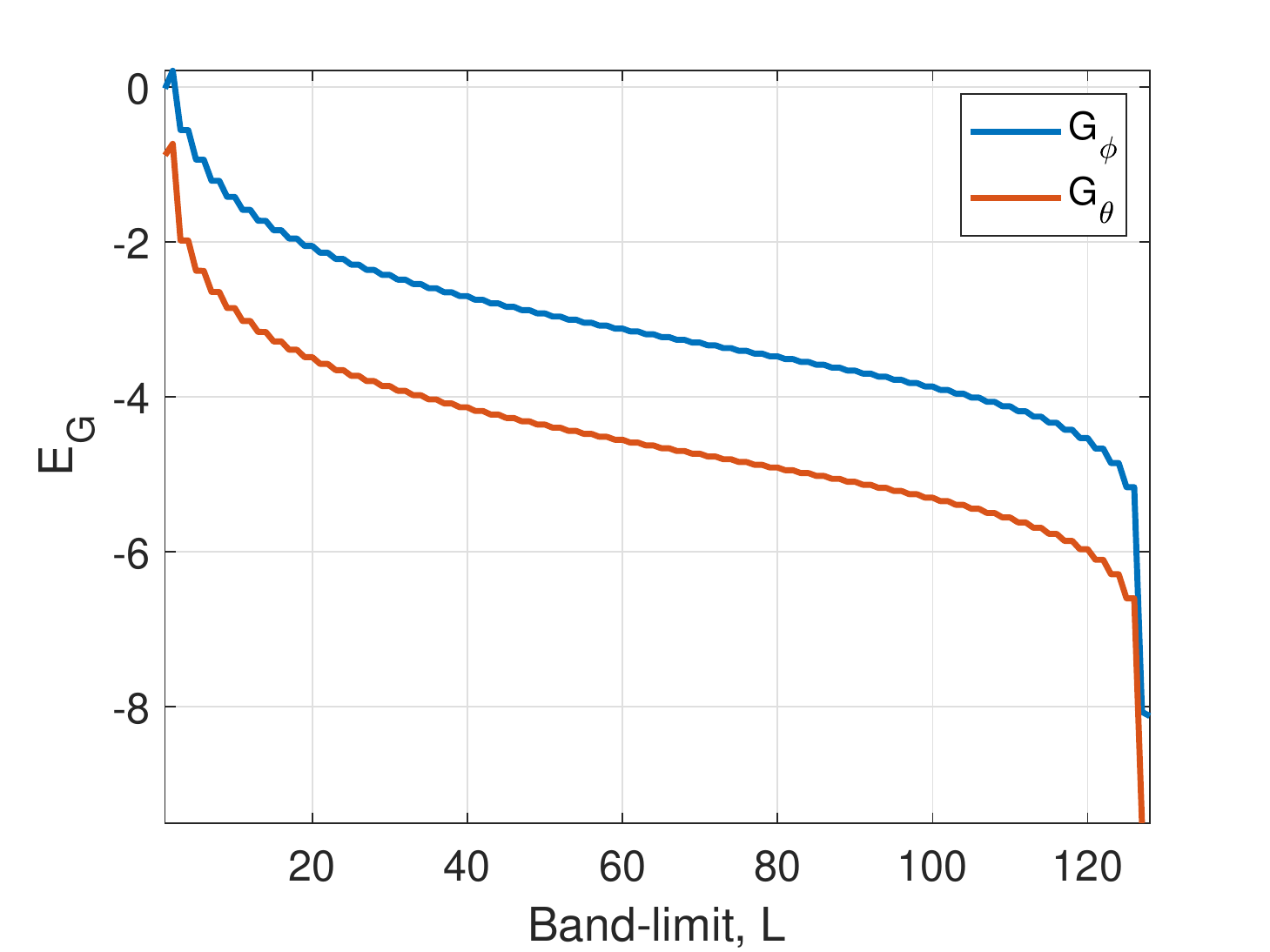}
\vspace{-1mm}
\caption{Reconstruction error $E_G$ of $\tilde{G}(\unit{x})$ for band-limit, $L$ from 1 to 128 shown in the logarithmic scale as $\log_{10}|\tilde{G}(\unit{x})|$}.
\vspace{-0mm}
\label{fig:gband}
\end{figure}
\begin{figure}[!t]
\centering
\vspace{-4mm}
\includegraphics[width=0.4\textwidth]{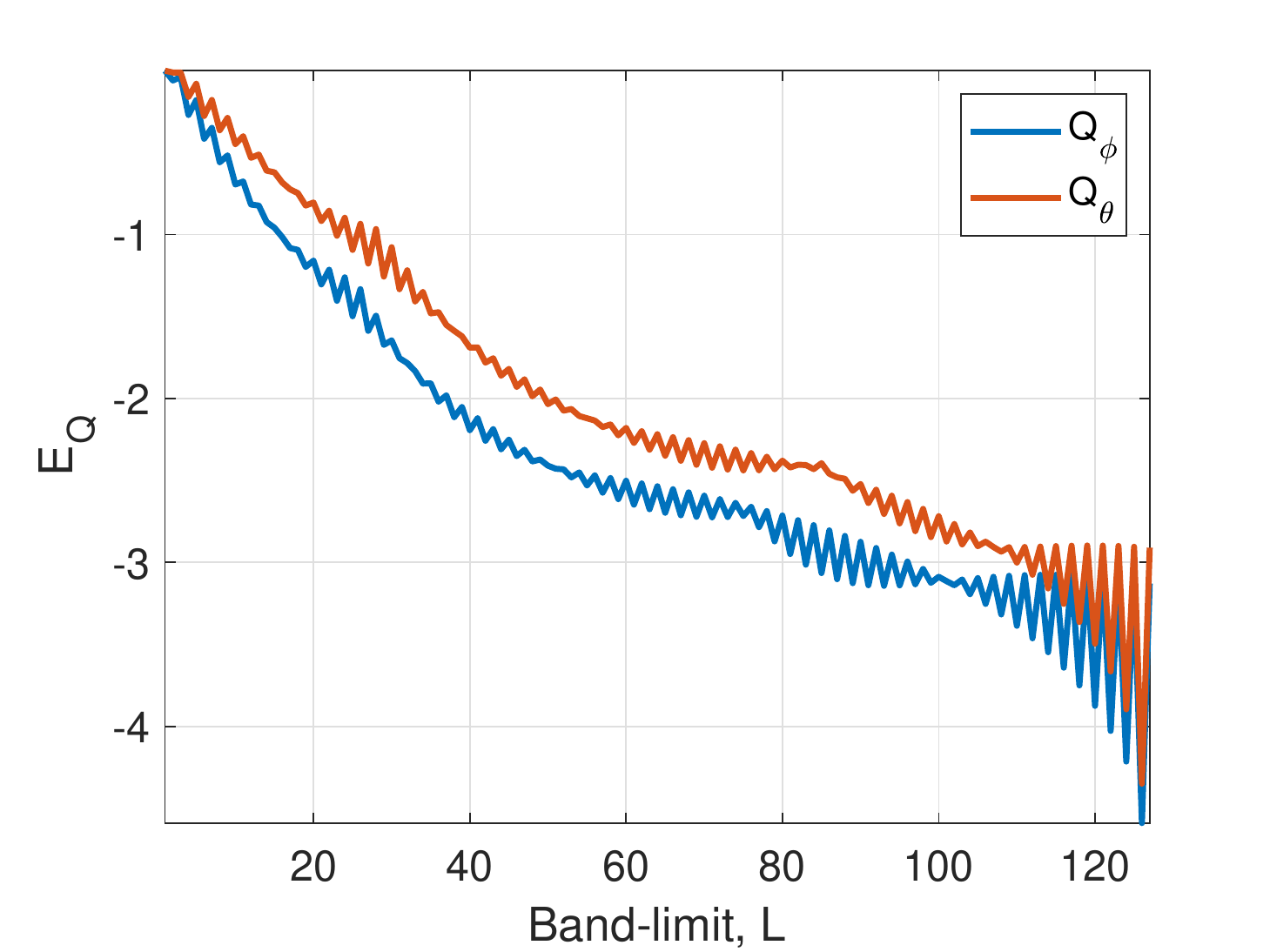}
\vspace{-1mm}
\caption{Reconstruction error $E_Q$ of $\tilde{Q}(\unit{x})$ for band-limit, $L$ from 1 to 128 shown in the logarithmic scale as $\log_{10}|\tilde{Q}(\unit{x})|$}.
\vspace{-0mm}
\label{fig:qband}
\end{figure}

\begin{figure*}[t]
    \vspace{-6mm}
    \centering
    \subfloat[$\tilde{G}_\theta$]{
        \includegraphics[width=0.2\textwidth]{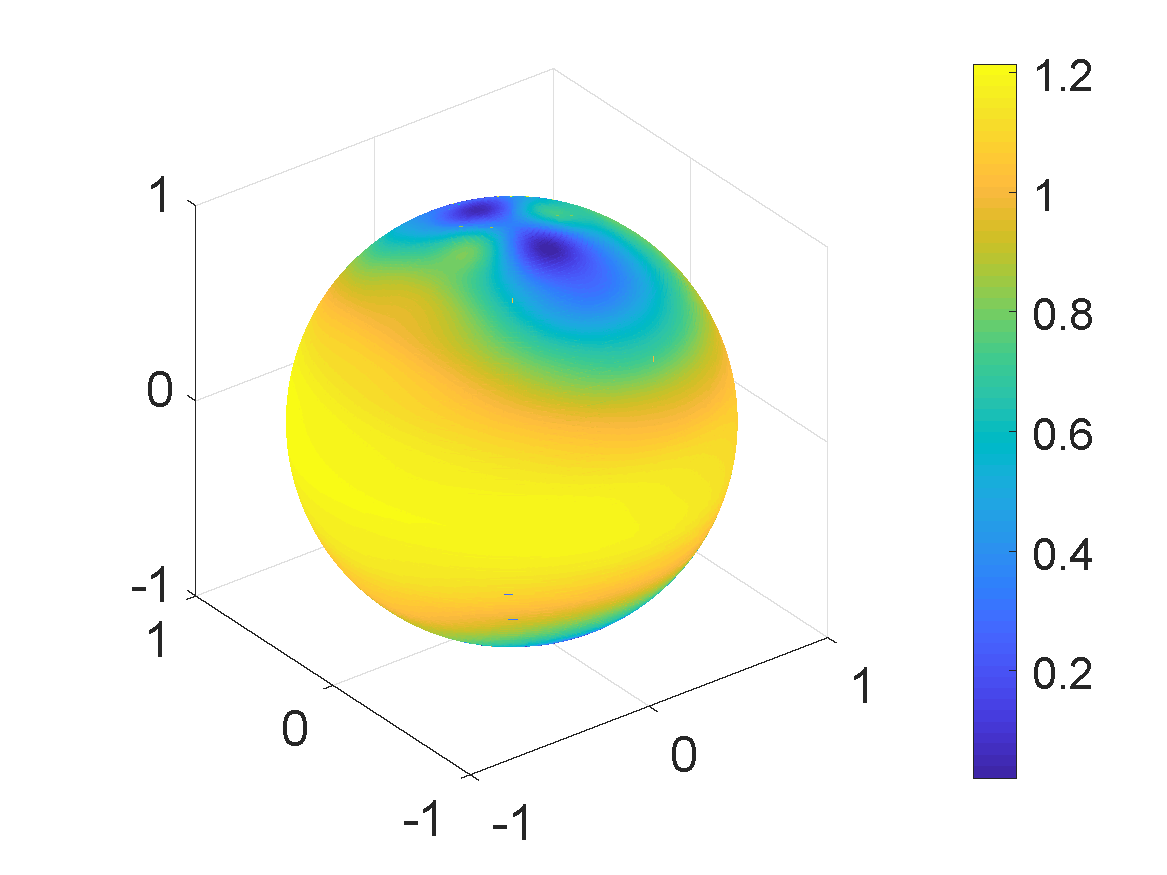}}
    \subfloat[$\tilde{G}_\phi$]{
        \includegraphics[width=0.2\textwidth]{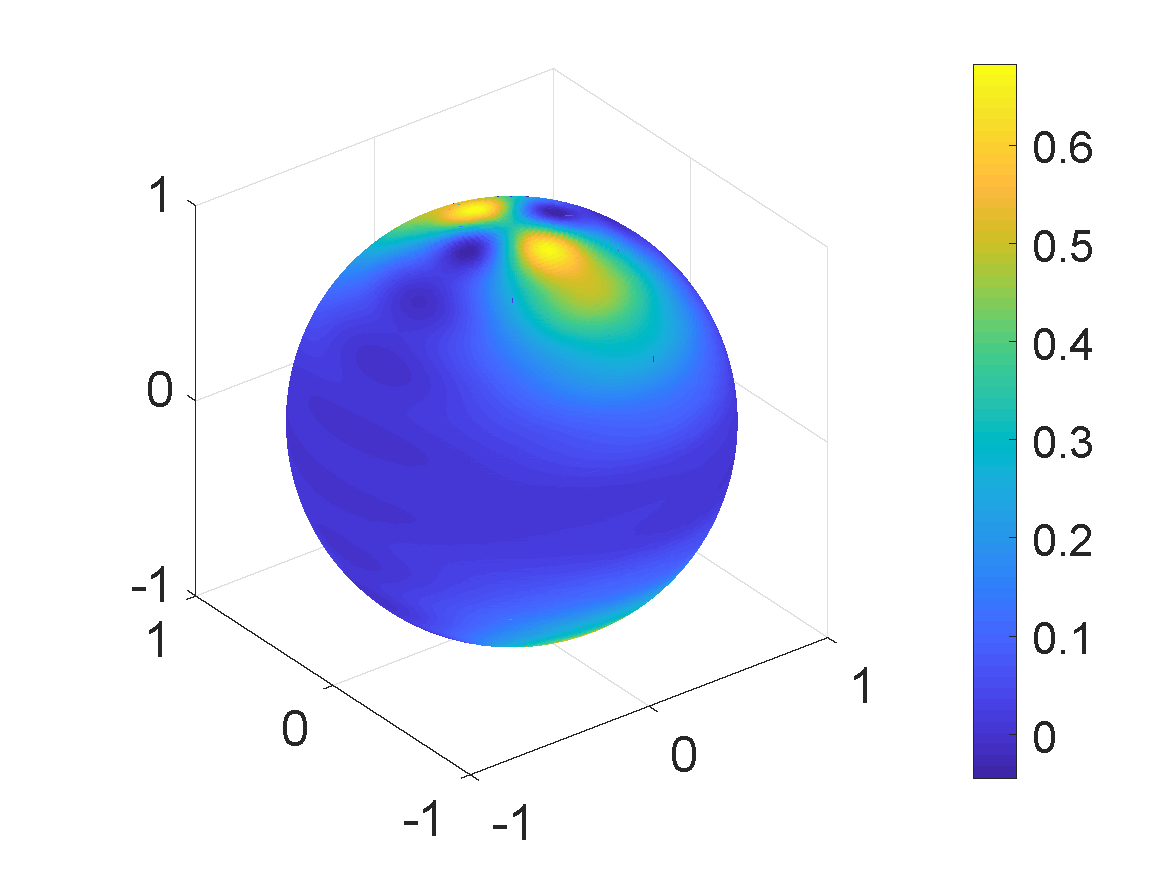}}
    \subfloat[$\tilde{Q}_\theta$]{
        \includegraphics[width=0.2\textwidth]{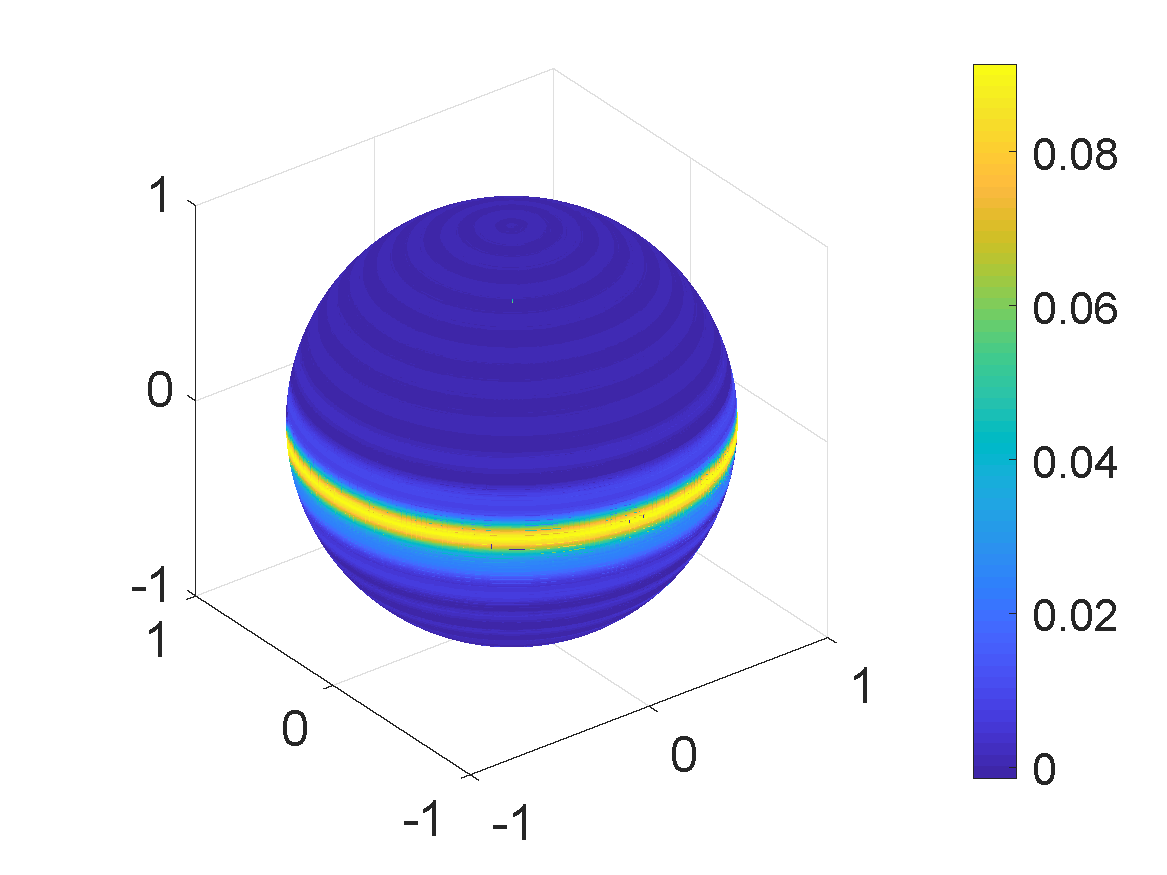}}
    \subfloat[$\tilde{Q}_\phi$]{
        \includegraphics[width=0.2\textwidth]{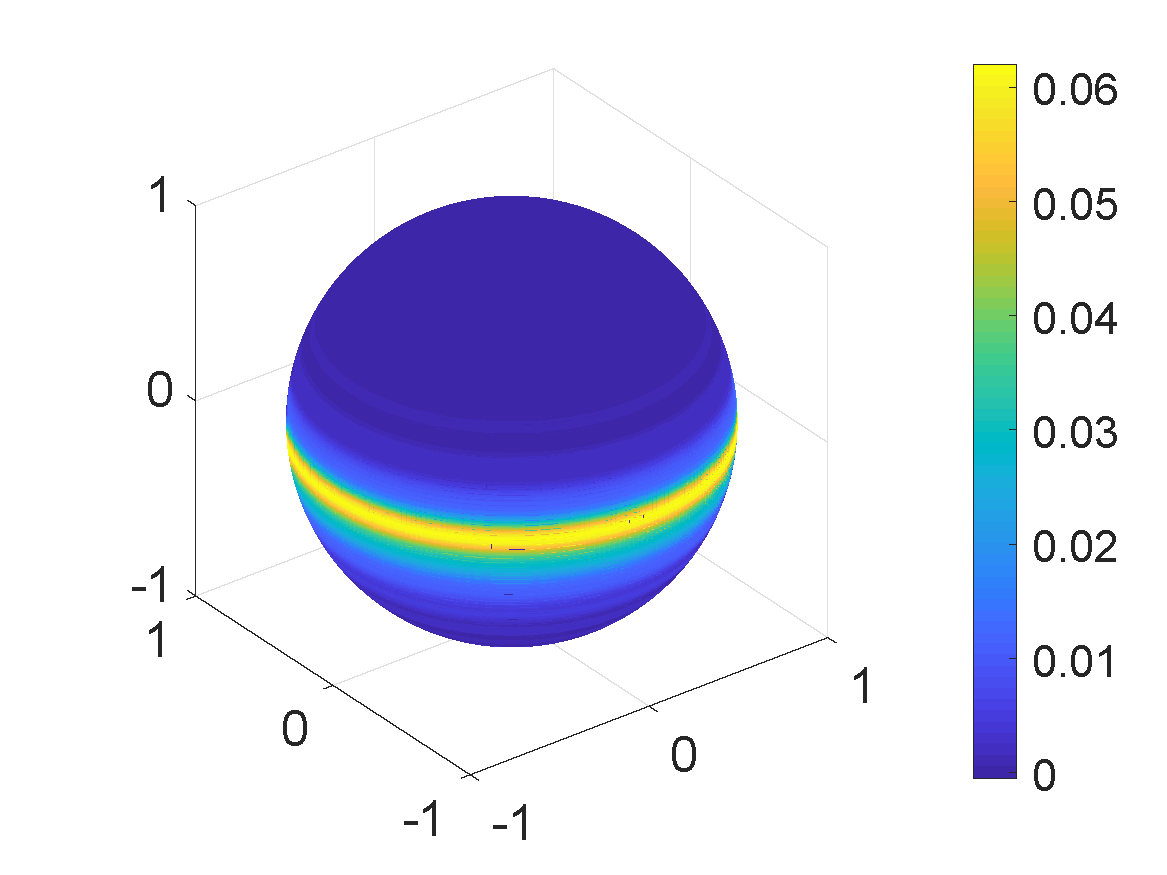}}
\caption{Reconstructed signals in spatial domain, band-limited at $L=69$. }
\vspace{-2mm}
\label{fig:spatial_plots}
\end{figure*}

\subsection{Equiangular Sampling}
In \cite{McEwen:2011}, equiangular sampling scheme has been proposed to compute the SHT of the signal band-limited at $L$. The equiangular~(EQ) scheme, denoted by $\sset{L}$, composed of $L+1$ rings of samples with equiangular spacing between the rings along the co-latitude along with one ring at the pole. In each iso-latitude ring, $2L-1$ equiangular samples are taken along longitude. The total number of samples in equiangular sampling are $L(2L-1) + 1 \sim 2L^2$. We need $2L-1$ samples in each ring for the reconstruction of the antenna radiation pattern. However, we only need $L+1$ samples along longitude in each ring for the computation of integral of the antenna radiation pattern. We define an equiangular scheme, denoted by $\sseq{L}$, that takes total of $(L+1)(L-1)+1 = L^2$ samples on $L$ rings along co-latitude with one ring at the pole and $L+1$ samples along longitude in each ring, for the exact computation of integrals involved in the evaluation of performance parameters provided the integrand is band-limited at degree $L$.

\subsection{Gauss-Legendre Sampling}
As an alternative to EQ sampling scheme, Gauss-Legendre (GL) sampling~\cite{Doroshkevich:2005}, based on Gauss-Legendre quadrature can be used for the computation of SHT and evaluation of integrals. Denoted by $\sglt{L}$, GL scheme takes $L$ rings of samples along the colatitude according to the Gauss-Legendre quadrature and $2L-1$ equiangular samples along the longitude in each iso-latitude ring. The total number of samples in this sampling scheme are equal to $L(2L-1) \sim 2L^2$. We also define $\sglq{L}$ that takes $L$ rings along co-latitude according to GL quadrature with $L+1$ samples along longitude in each ring.

\subsection{Optimal Dimensionality Sampling}
Both GL and EQ sampling strategies require samples on the order of $2L^2$ for the computation of SHT and reconstruction of radiation pattern and $L^2$ samples for the evaluation of quadrature. Recently, another iso-latitude sampling method, referred to as optimal dimensionality~(OD) sampling scheme, has been proposed in \cite{Khalid:2014} that takes $L^2$ samples for both the SHT computation and quadrature evaluation. This sampling scheme, denoted by $\sso{L}$, takes samples on the $L$ iso-latitude rings with $1,3,5,\hdots,2L-1$ samples along $\phi$ in each ring.

\subsection{Antenna Pattern Reconstruction and Computation of Performance Parameters}
The band-limit of the radiation pattern and the incoming signal average power defined by HUT model can be determined employing \eqref{Eq:tildeG} and \eqref{Eq:error_bound} for given value of $\epsilon_o$. We assume that the radiation pattern $G$ and the incoming signal average power~(either $\theta$ or $\phi$ polarization) $Q$ are band-limited at spherical harmonic degrees $L_G$ and $L_Q$ respectively. Utilizing any of the GL, EQ and OD sampling for the acquisition of measurements of radiation pattern and incoming signal average power, the two signals can be reconstructed with control on the norm of reconstruction error.  We note that the product of radiation pattern and incoming power signal distribution is band-limited at $L_G+L_Q-1$. For the reconstruction of antenna radiation pattern $G$, we can use  any of the $\sset{L_G}$, $\sglt{L_G}$ and $\sso{L_G}$ sampling schemes which respectively take (asymptotically) $2L_G^2$, $2L_G^2$ and $L_G^2$ samples on the sphere. For the evaluation of directivity or MEG, any of the $\sseq{L_G}$ or $\sseq{L_G+L_Q-1}$, $\sglq{L_G}$ or $\sglq{L_G+L_Q-1}$ and $\sso{L_G}$ or $\sglq{L_G+L_Q-1}$ sampling schemes, each of which take $L_G^2$ or $(L_G+L_Q-1)^2$ samples for quadrature calculation, can be used to calculate the required integrals. Although GL and EQ sampling schemes enable accurate computation of SHT, these schemes may become impractical for large band-limits due to the dense sampling around the poles. The optimal dimensionality~(OD) sampling scheme in contrast is more practical as 1) it requires half the number of samples for the computation of SHT, 2) same number of samples for the evaluation of quadrature and 3) the samples are more uniformly distributed on the sphere. We analyse the band-limit of antenna radiation pattern and incoming signal power distribution and evaluate the MEG using the sampling schemes as proposed here in the next section.


\vspace{-1mm}
\section{Analysis}
Here we analyse the accuracy of the reconstruction of antenna radiation pattern of a PIFA type antenna and the incoming signal power distribution. The radiation pattern for PIFA is obtained at the center frequency of GSM ($f_c = 1842$ MHz) from MATLAB antenna toolbox \footnote{MATLAB uses the Method of Moments (MoM) to generate antenna patterns.}. For different values of band-limits $1\le L \le 128$, we determine the error $E_G$ defined in \eqref{Eq:error_bound} employing equiangular sampling scheme $\sset{L_g}$, which is plotted in \figref{fig:gband}, where it is evident that the $E_G$ decays to zero with the increase in band-limit. We choose the band-limit $L_G=20$ such that $E_G<\epsilon_o=10^{-2}$ and plot the reconstructed antenna radiation $\tilde G$ pattern in \figref{fig:spatial_plots}(a) and \figref{fig:spatial_plots}(b). We also use the same method to determine $L_{Q_{\theta}}=L_{Q_{\phi}}=50$ for HUT model of the incoming signal given in \eqref{Eq:hut} corresponding to parameters averaged over all environmental conditions such that $\theta_o=1.6^\circ$, $\sigma^{-}=5.5^\circ$ and $\sigma^+ =8.6^\circ$ for $Q_\theta$ and $\theta_o=1.8^\circ$, $\sigma^{-}=7.4^\circ$ and $\sigma^+ =13.7^\circ$ for $Q_\phi$. We plot the reconstruction error $E_Q$ in \figref{fig:qband} and reconstructed $\tilde Q_\theta$ in \figref{fig:spatial_plots}(c) and \figref{fig:spatial_plots}(d).

\begin{figure}[!t]
\centering
\vspace{-3mm}
\includegraphics[width=0.478\textwidth]{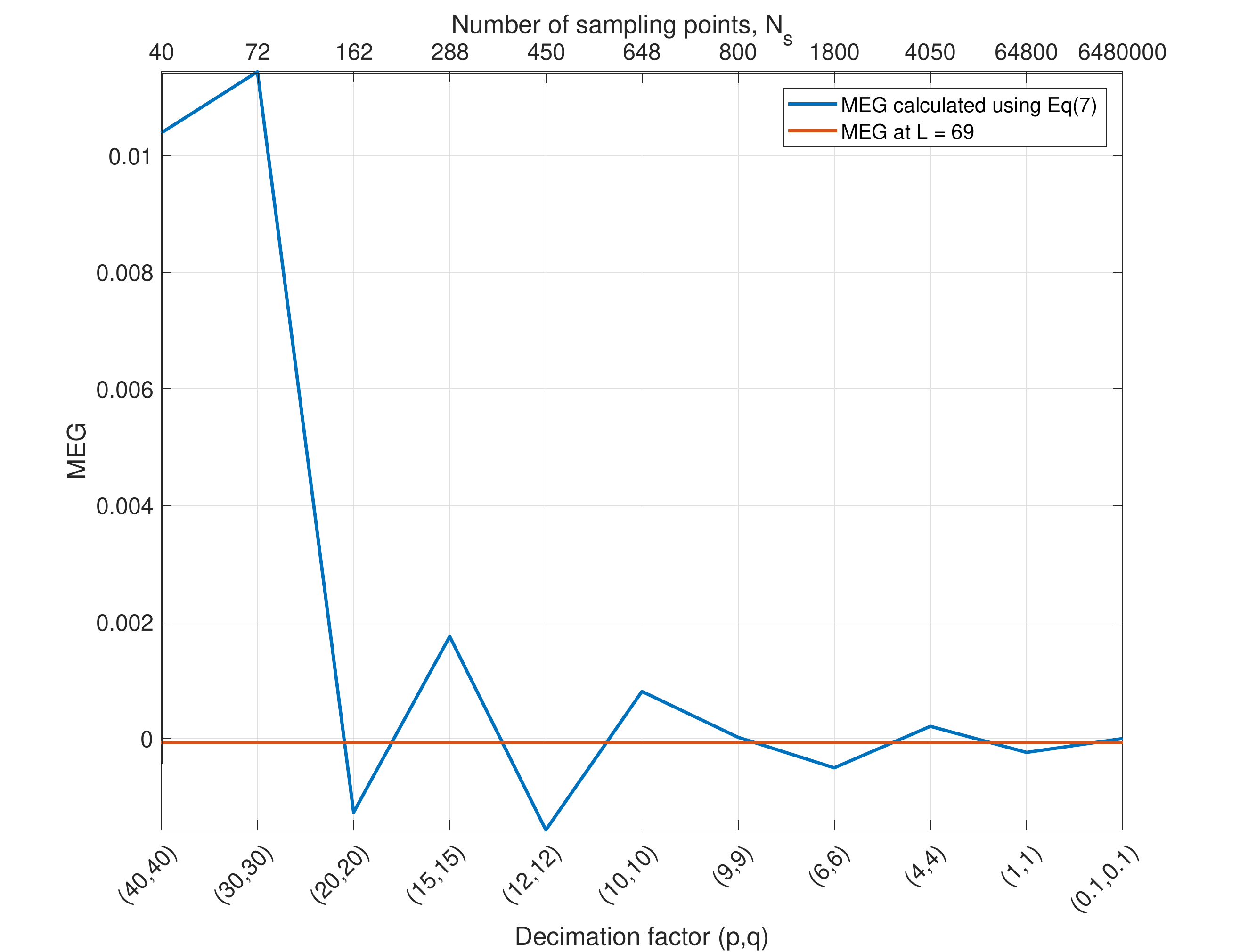}
\vspace{-1mm}
\caption{Mean effective gain (MEG) calculated using \eqref{Gamma_sampling_method} with HUT power model, as a function of decimation factors $(p,q)$. The solid red line shows the MEG calculated using $\sseq{L_G+L_Q-1}$ quadratures for $L_G = 20$ and $L_Q = 50$. All the values are normalized to MEG with $(p,q)=(0.1^\circ,0.1^\circ)$ and shown in the logarithmic scale as $\log_{10}|{\rm MEG}|$.}
\vspace{-3mm}
\label{fig:meg}
\end{figure}

We also evaluate the MEG $\Gamma$ by taking measurements on $\sseq{L_G+L_Q-1}$ sampling grid for the incoming signal power model and PIFA antenna radiation pattern. For comparison, MEG is also calculated by using the formulation in \eqref{Gamma_sampling_method} for different decimation factors for sampling of points along $\theta$ and $\phi$. From \figref{fig:meg} it is evident that, in order to achieve the same accuracy in the computation of MEG, for a decimation factor of $(p,q) = (0.1^\circ,0.1^\circ)$ corresponding to $Ns=6,480,000$ points, the proposed schemes requires only $L^2 = (70)^2 = 4900$ points in total.

\vspace{-1mm}
\section{Conclusions}

In this paper, we have proposed the use of Equiangular, Gauss-Legendre and  optimal dimensionality schemes to sample and reconstruct the spherical antenna radiation patterns as well as the incoming signal power for a real-world model of power distribution. We have developed a formulation to appropriately choose the sampling grid and the number of measurements on the chosen grid for the reconstruction of antenna radiation pattern and computation of its performance parameters such as directivity and mean effective gain~(MEG). Using our methodology, we have showed that MEG can be calculated with the same order of accuracy by taking significantly less number of points than required by the existing schemes that support the computation of MEG using numerical quadrature.

\bibliographystyle{IEEEbib}
\bibliography{IEEEabrv,refs}

\end{document}